# The Ambivalence of Cultural Homophily:

**Field Positions, Semantic Similarities, and Social Network Ties in Creative Collectives**

Nikita Basov

*Centre for German and European Studies*

*St. Petersburg State University*

**Abstract**

This paper utilizes a mixture of qualitative, formal, and statistical socio-semantic network analyses to examine how cultural homophily works when field logic meets practice. On the one hand, because individuals in similar field positions are also imposed with similar cultural orientations, cultural homophily reproduces 'objective' field structure in intersubjective social network ties. On the other hand, fields are operative in practice and to accomplish pragmatic goals individuals who occupy different field positions often join in groups, creatively reinterpret the field-imposed cultural orientations, and produce cultural similarities alternative to the position-specific ones. Drawing on these emergent similarities, the cultural homophily mechanism might stimulate social network ties between members who occupy not the same but different field positions, thus contesting fields. I examine this ambivalent role of cultural homophily in two creative collectives, each embracing members positioned closer to the opposite poles of the field of cultural production. I find different types of cultural similarities to affect different types of social network ties within and between the field positions: Similarity of vocabularies stimulates friendship and collaboration ties within positions, thus reproducing the field, while affiliation with the same cultural structures stimulates collaboration ties between positions, thus contesting the field. The latter effect is visible under statistical analysis of ethnographic data, but easy to oversee in qualitative analysis of texts because informants tend to flag conformity to their positions in their explicit statements. This highlights the importance of mixed socio-semantic network analysis, both sensitive to the local context and capable of unveiling the mechanisms underlying the interplay between the cultural and the social.

# 1. Introduction

In accordance with the mechanism of homophily (Cohen, 1978; Kandel, 1978; McPherson, Smith-Lovin and Cook, 2001; Dahlander and McFarland, 2013), 'networks and culture' research increasingly reports a positive effect of cultural similarity on social network ties (DiMaggio, 1987a; Mark, 1998; Lizardo, 2006; Vaisey and Lizardo, 2010)[1]. Given that individuals similarly positioned in social fields tend to be imposed with similar cultural orientations (Bourdieu, 1993; 1996), in light of Bourdieu's field theory cultural homophily is reproducing field structures of 'objective relations' between social positions in social networks of 'intersubjective relations' between the individuals who occupy these positions (Bourdieu and Wacquant, 1992; De Nooy, 2003; Bottero, 2009; Bottero and Crossley, 2011).

Simultaneously, there are good reasons to expect practice to refract the effect of field-imposed cultural similarities on social network ties. Engaging in practice, individuals often creatively reinterpret field-imposed cultural orientations with respect to tasks at hand (Bourdieu, 1984, 1990, 1996a) and small groups of joint practice often bring together individuals who occupy different field positions. This may lead to the transformation of field-imposed cultural orientations (De Nooy, 2003) and to the emergence of local group cultures across positions (Basov, De Nooy and Nenko, 2018). Drawing on local cultural similarities emerging in groups, cultural homophily may stimulate and reinforce social ties between different field positions, thus contesting objective relations in addition to reproducing them.

My paper investigates this ambivalent nature of the cultural homophily mechanism, asking: How does cultural homophily work in small groups of shared practice that bring together different field positions?

To address this question, I take the increasingly popular social and cultural duality perspective (Mohr and Duquenne, 1997; Breiger, 2000; Mische and Pattison, 2000; Fuhse, 2009; Godart and White, 2010; Basov and Brennecke, 2017) which responds to the criticisms of social network analysis for treating culture as secondary to social network structure (e.g., Emirbayer and Goodwin, 1994; Vaisey and Lizardo, 2010) and is aligned with the overall 'cultural turn' in sociology (Friedland and Mohr, 2004). While sociology has long made a distinction between (social) structure and culture (e.g., Parsons and Shils, 1951; Sewell Jr, 1992), the duality perspective pursues a structural account not only of the social but also of the cultural (Mohr, 1998). More specifically, this perspective views culture as distinct and self-consistent "continuously interacting populations of forms of discourse" (White and Godart, 2007: 11), composed of "relations rather individual elements and the patterns arising in these" (Edelmann and Mohr, 2018: 3), embedded in specific constellations of social network ties but having an 'equal weight' to them (Mohr, 1998). In line with this perspective, I examine the mechanism of cultural homophily as the effect of being affiliated with the same cultural structures on social network ties.

[1] The latter paper even finds poor inverse effect of social networks on cultural similarities when cultural homophily is controlled for.

To observe how cultural homophily works in between field and practice, I study two creative collectives operating in the field of cultural production (see also Pivovarov and Nikiforova, 2016; Nenko, 2017). Each of the collectives brings members who occupy opposite positions in this field to interact and engage in joint practice. Such an empirical setting enables both culture imposed by the members' 'objective' positions in the field and local group culture (see Basov et al., 2018) and, hence, allows for the presumed ambivalence of cultural homophily.

In order to account not only for the field-imposed culture, but also for the local culture emerging from practice (see Bourdieu, 1990; Mohr and Neely, 2009: 220; Basov et al., 2018), this paper aligns with the rapidly expanding mixed method movement in network analysis (Fuhse and Mützel, 2011; Bellotti, 2014; Domínguez and Hollstein, 2014; Bellotti, 2016; Ibrahim and Crossley, 2016; Basov, 2018). It does so in at least two ways. On the one hand, I use a combination of qualitative (interviews, written texts, interaction observations) and quantitative (sociometric survey) network data sources. On the other hand, I blend qualitative and quantitative analytical techniques (see Crossley and Edwards, 2016) in a mixed-method inductive version (see also Christopoulos and Vogl, 2015; Basov et al., 2018) of socio-semantic network analysis (Roth and Cointet, 2010; Basov and Brennecke, 2017; Hellsten and Leydesdorff, 2017; Saint-Charles and Mongeau, 2018). Firstly, I inductively map semantic networks as stable word associations in a multisource textual corpus of members' expressions coming from ethnographic studies of the collectives. This makes a contrast to focusing on the 'broad orientations' derived from large-scale surveys that currently prevail in the studies of cultural homophily (e.g., Mark, 1998; Lizardo, 2006; Vaisey and Lizardo, 2010) but are continuously criticized for being insensitive to local practical contexts (Fine and Kleinman, 1979: 5-6, 17-18; Fuhse, 2009: 66; Breiger and Puetz, 2015: 559). Multidimensional scaling is applied to the mapped semantic networks to preliminary check how culturally similar individuals are in the same and in different field positions, within and across the collectives.

Secondly, I draw on the 'full' version (Lee and Martin, 2018) (see also Godart and Galunic, 2019) of the two-mode approach (Breiger, 1974) and relate semantic networks of the collectives to their social networks using a two-layer socio-semantic framework, where two-mode links connect words to individuals who use them. Hence, the cultural and the social are treated as dually positioned structures of a different nature, both interrelated and self-organizing. To infer the mechanism of cultural homophily, I statistically examine these two-layer socio-semantic networks with the help of recently developed MERGMs: multilevel exponential random graph models (Wang, Robins, Pattison and Lazega, 2013). MERGMs control for structuring mechanisms specific for each of the subnetworks (here, social, semantic, and the two-mode network between them) while modeling higher-order mechanisms that relate these subnetworks. It enables controlling for other social and socio-cultural structuring mechanisms while examining the mechanism of cultural homophily.

Finally, I consider the particular words and associations that, according to the models, affect certain types of social ties within and between field positions. I go back to the qualitative data to examine usage of these words and associations by particular friends and collaborators in similar and in different positions. In doing so, I aim to understand these words and associations in their local socio-cultural context (for a more extensive argumentation of this approach and an application, see Basov et al., 2018).

The mixed approach applied in this paper is intended to be both sensitive to the context and able to 'totalize' across practical situations (Bourdieu, 1990: 82), finding the patterns beyond those captured by an ethnographer's eye with the help of statistical analysis (for similar arguments see De Nooy, 2009; Crossley and Edwards, 2016).

The results contribute both to our understanding of the relationship between the social and the cultural in social fields and to the methodology for the analysis of this relationship.

# 2. Cultural Homophily Between Field and Practice

According to Pierre Bourdieu, individuals are positioned in social space with regard to each other both objectively, based on the types and volumes of capital they have access to, and subjectively, based on how the agents interpret their own and others' positions. Jointly, objective and subjective positioning shape social fields—relational systems, where the competition of individuals for different forms of capital unfolds. These relational systems are specific to different areas of social life, such as art, academia, and religion, and each field is formed by a field-specific set of objective relations and shared subjective understandings of what goes on in the field (Bourdieu, 1984; Bourdieu and Johnson, 1993; Bourdieu, 1996a).

There are also relations between social properties held by agents in certain positions and respective categories of perceptions (high/low, masculine/feminine, large/small, etc.). These categories—for instance, through classification (De Nooy, 1999, 2003; Childress, Rawlings and Moeran, 2017)—form cultural categorizations attached to individuals who occupy certain social positions. Owing to a common position-specific habitus, i.e. "durable, transposable dispositions" (Bourdieu, 1990: 53), members of a field occupying similar positions and hence having similar economic statuses, educational backgrounds, and occupations, tend to have similar cultural orientations corresponding to their positions, therefore having similar perspectives on what is good and what is bad, who is strong and who is weak, and so on; they also exhibit similar cultural tastes and preferences (Bourdieu, 1984; DiMaggio, 1987a; Erickson, 1996; Lizardo, 2006). For example, intellectuals are known to have similar preferences in cuisine (Bourdieu, 1996a), arts, and photography (Bourdieu, 1984).

Although Bourdieu famously criticizes social network analysis for substituting examination of objective relations that invisibly structure society with analysis of visible interactions, hence taking effects (social ties) for causes (objective relations), social networks are not irrelevant to his field theory. In fact, Bourdieu explicitly says that field structure "is different from the more or less lasting networks through which it manifests itself" (Bourdieu and Wacquant, 1992: 113-114), and, therefore, while it is important to distinguish between the two, both are present, and social networks are an explication of objective relations. Bourdieu also uses the notion of network in his definition of social capital as ''the sum of the resources, actual or virtual, that accrue to an individual or a group by virtue of possessing a durable network of more or less institutionalized relationships of mutual acquaintance and recognition'' (Bourdieu and Wacquant, 1992: 119).

Subsequent literature further highlights the importance of social networks for Bourdieu's field theory (De Nooy, 2003; Bottero, 2009; Bottero and Crossley, 2011), arguing that "[w]ithout an account of relationships and networks, actors in Bourdieu's model are atomized and he lacks an account of the mechanisms which generate similarities in their habitus" (Bottero and Crossley, 2011: 101). In addition, network analysis proves helpful to test Bourdieu's theorizing empirically (e.g., DiMaggio, 1987a; Erickson, 1988, 1996; De Nooy, 1999; Lizardo, 2006; Breiger and Puetz, 2015). In particular, based on Bourdieu's argument that individuals occupying similar positions in social fields tend to be imposed with similar cultural orientations, which are translated into intersubjective relations between them (Bourdieu and Johnson, 1993; Bourdieu, 1996a), researchers highlight that "actors are more likely to form and sustain social contacts with those socially similar to themselves" (Bottero and Crossley, 2011: 102). Scholars also argue that "taste is <…> a means of construction social relations <…> It helps to establish networks of trusting relations that facilitate group mobilization" (DiMaggio, 1987a: 443), and that sharing of cultural tastes and worldviews stimulates social network ties (Lizardo, 2006; Vaisey and Lizardo, 2010).

Translation of position-imposed cultural similarities into social network ties implies cultural homophily, a mechanism that stimulates ties between individuals with similar cultural orientations (Cohen, 1978; Kandel, 1978; DiMaggio, 1987a; Mark, 1998; Lizardo, 2006; Vaisey and Lizardo, 2010; Dahlander and McFarland, 2013; Basov and Brennecke, 2017), and, therefore, serves the reproduction of objective relations at the intersubjective level, as described by Bourdieu.

The perspective on cultural homophily as serving the reproduction of objective relations in social networks is incomplete without the other side of Bourdieu's theorizing: his theory of practice. According to Bourdieu, objective relations are not merely mirrored by intersubjective relations. The latter are characterized by a struggle for better field positions with corresponding capitals and, simultaneously, for cultural categorizations relating field positions to each other. Consequently, when objective relations are put to real practice, they are intersubjectively adjusted with respect to the 'matters at hand' (Bourdieu, 1984, 1990, 1996a). Moreover, to accomplish complex tasks by combining diverse forms of capital, practice often brings together individuals occupying different field positions. Interacting in joint practice, these individuals get to bear each other's cultural orientations (De Nooy, 2003: 323). As a result, "interpersonal relations mediate and transform the effect of objective relations" (De Nooy, 2003: 305). This enables emergent culture that reconfigures field-imposed cultural categories (Basov et al., 2018).

If one agrees with Bourdieu and De Nooy, she is to admit that in local practical contexts, e.g., in small groups of joint practice that are known to produce their specific cultures (Fine, 1979), the mechanism of cultural homophily may lead to the establishment of intersubjective ties not only between individuals who occupy similar positions in social fields, but also to ties between different positions, thus contesting field structure. Then, understanding of the culture-structure relationship in social fields requires shedding light on how cultural homophily works in local practical contexts that combine different positions. There, the field forces, which tear the social network of the group apart by imposing position-specific cultures, meet practice, which connects the group with the emergent culture shared across positions.

Homophily is shown to work for collaboration networks (McPherson et al., 2001; Dahlander and McFarland, 2013) and for friendship networks (Kandel, 1978; Mark, 1998; Lizardo, 2006; Vaisey and Lizardo, 2010). When analyzing a professional field, such as the field of cultural production studied here, it is, perhaps, obvious to consider collaborations, such as joint creation of artworks, preparation of exhibitions and presentations, project coordination, or fundraising. Indeed, art is known as a collective endeavor pursued by collaborative circles (Farrell, 2003), where working relationships rely on similarities in views (Bourdieu and Johnson, 1993), and changes in artistic paradigms are shown to rely on coordinated action combined with coherence in styles (White, 1993). At the same time, the quest for novelty characteristic of creative work is emotionally challenging. Artists were documented to seek emotional support from those with similar views and maintain friendship ties with them (Farrell, 2003; Elsbach, 2009). Furthermore, differences in how culture interplays with friendship ties and with collaboration ties between artists are empirically captured (Basov and Brennecke, 2017). It makes sense, therefore, to compare how cultural homophily works with respect to friendship ties and to collaboration ties.

# 3. The Opposite Positions in the Field of Cultural Production

According to Bourdieu and Johnson (1993), the field of cultural production historically exists between two poles guided by two opposing principles of heirarchization: autonomous and heteronomous. The latter is strongly affected by financial imperatives of the economic sphere, oriented at profit, and hence is aimed at broad audiences. It involves large-scale production of the so-called low-brow—mass or popular—culture, sustained by a vast and complicated culture industry. Success at this pole of the field depends on the economic capital. The autonomous pole is the opposite to it, driven by the field's internal logic of heirarchization that relies on symbolic capital and restricted production of high-brow culture for the narrow audiences sufficiently trained to consume its goods. It is the autonomous pole that allows the field of cultural production to historically develop a degree of autonomy from other fields, especially, the field of power.

Positioned closer to one of the two field poles, individuals compete for corresponding forms of capital and have different cultural orientations, summarized by DiMaggio (1987b: 83) as "aesthetic versus managerial orientation". The logic of autonomous cultural production stimulates creatives to seek novel modes of representation in order to achieve symbolic recognition (Crane, 1989; Bourdieu and Johnson, 1993; Giuffre, 1999). Individuals in these positions value freedom, creativity, originality, individualism, self-expression over pragmatism, and the process of creation over the result. In contrast, individuals positioned closer to the heteronomous pole of the field follow a different logic, the one more similar to the logic of business management which they 'smuggle' into arts acting as 'double personages' (Bourdieu, 1996: 216), sharing utilitarian aspirations and valuing market success among broad publics (Bendixen, 2000; DeVereaux, 2009). Typical orientations are towards project thinking, teamwork, economic and operational effectiveness, control, status, and broad networking with other agents of the field. Individuals in these positions are usually marketing-, management-, and service-oriented. Pragmatism is preferred over expression, market result—over creative process.

While those positioned closer to the autonomous pole of the field, or ‘auctors’, are known to opt for the relatively prestigious but risky and precarious careers of independent artists, those closer to the heteronomous pole, or ‘lectors’, stick to the less ambitious but safer careers of teaching and cultural management (Bourdieu and Johnson, 1993). For the sake of simplicity, further I label the former ‘artists’ and the latter—‘managers’. However, one should not mistake the autonomy- and heteronomy-oriented social positions (and, especially, the labels I use) for functions or organizational positions. The former imply certain cultural orientations and aspirations, which condition preferences for the latter. In other words, individuals choose functions and organizational positions (and even organizations themselves) because of social positions, not vice versa. And when individuals have to take functions not typical for their social position, they feel and express discontent. For instance, when in the collectives I study some of the autonomy-oriented ‘artists’ who have to promote their work to a broader audience they are usually unhappy with having to take this function because their main values are still creative freedom, novelty, individualism, and self-expression and their main capital of interest is symbolic rather than economic. Moreover, quite often functions emerge because of social positions. For example, owing to the aspirations induced by their social position, the heteronomy-oriented ‘managers’ in one of the collectives I study here introduced educational activities to their collective.

With the recent expansion of ‘cultural industries’ (Hesmondhalgh, 2007) and ‘creative economy’ (Howkins, 2002; Kuleva, 2018) agendas and subsequent strengthening tendencies towards heteronomy in the artistic field (Cook, 2001; Hewison, 2014), the heteronomy-oriented positions increase in ‘distinctiveness’, up to complete separation of the arts management from the arts. Bendixen (2000: 4) highlights with the very first words of his programmatic article on arts management: “[w]hen we speak of arts management, we are speaking of management, not art” (for empirical evidence on opposing arts and arts management, see Kuesters, 2010).

At the same time, in creative economies artistic work requires more and more rigorous coordination and marketing efforts in order to gain attention and access not only the economic but also the symbolic capital autonomy-oriented individuals struggle for. So, despite the growing distinction, autonomous artists and cultural managers also have to cooperate more and more intensely and even join each other in creative organizations, where artistic and managerial labor can be divided. However, this cooperation is not an easy one to maintain, because of the inherent ‘natural’ conflict between orientations of the two positions imposed by the field and, moreover, a perception of supremacy over each other. Artists tend to view managers as the "'merchants in the temple' of art" (Bourdieu, 1996b: 216), satisfying low-brow tastes of the wider public for the sake of money, whereas managers, often governing artists' activities, see them as disorganized subordinates whom they need to control and artworks—as goods to sell. This, in turn, irritates the artists perceiving themselves as creators of symbolic values, who are beyond economic evaluation and the process of creation—as a sacred process, not subject to management (see also: Bendixen, 2000).

The case of artists and managers who meet in creative collectives is thus not only topical but also suiting my purposes. It involves both the inherent field-imposed cultural differences and the possibility for the emergence of cultural similarities throughout joint practice, hence, cultural homophily in such collectives may both reproduce and contest the field of cultural production.

# 4. The Two Positions in the Barcelona and Madrid Collectives

The Barcelona collective (Figure 1) was founded in 1993 as an artistic residence bringing together members with diverse backgrounds, artistic genres ranging from graphics, photography, and filming to design and installations. In 2003, the group moved into a new space downtown of Barcelona, designed to host educational, research, experimentation, and communication initiatives along with the artists' studios. In 2011, it launched educational programs for children and families aimed at developing creative thinking and artistic imagination and mostly consisting of workshops by members of the collective. In conducting these programs, the collective maintains a strong partnership with one of the major foundations for contemporary art in Barcelona. It is also supported by commercial companies with recognized brands. As a rule, artists work individually, producing material pieces intended for exhibitions. However, some of the members team-up to do joint creative projects. Besides, each of the member artists is to participate in the educational workshops for children, conducted by the collective. Additionally, the collective holds open door days, when all the eleven core members participate: Some present their works, some give lectures, and some run master-classes, often cooperating with each other.

The Madrid collective (Figure 2) was created in 2011 by a group of architects who engaged in institutional critique and alternative architecture practices. With the goal of stimulating public discussion and reflection on the production and reuse of trash and waste, they took off by organizing a series of corresponding events. Now, the nine core members work together to (re-

)appropriate the alienated public spaces for citizens using a number of creative formats, tools, and partnerships, and utilize recycling materials, such as plastic, old tires, and wood. The main artistic activities are installations and public space interventions, accompanied by workshops, conferences, festivals, curator projects, and exhibitions. Aiming to re-inhabit city spaces together with local citizens, the collective follows a participatory approach. All the projects are open for access and participation and involve local publics—citizens and NGO representatives. The activities take place in some of the most visible creative spaces in Madrid and extend worldwide.

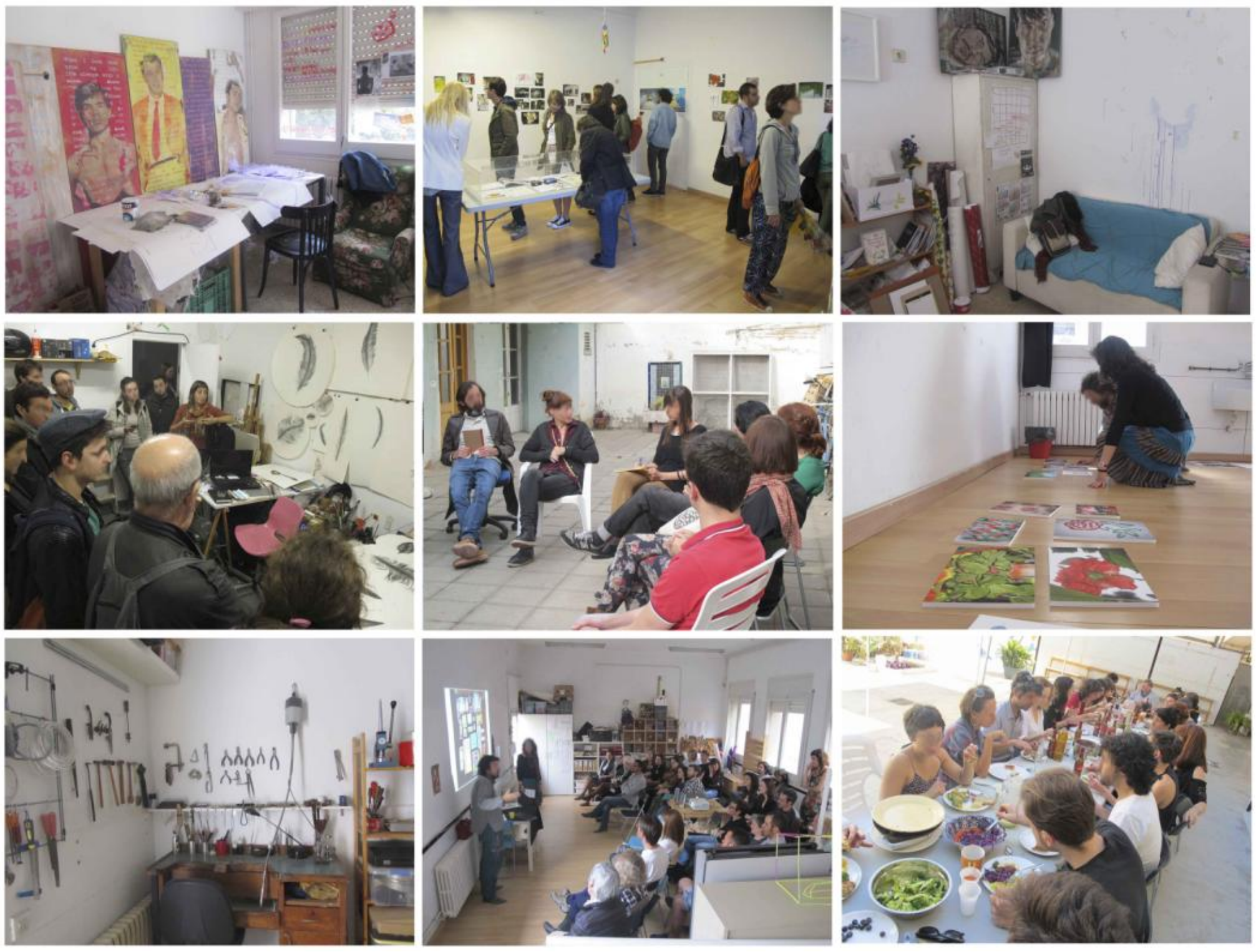

**Figure 1**. The Barcelona collective.

Both autonomy-oriented 'artists' and heteronomy-oriented 'managers' are present in each of the collectives. The former mainly value individual creative experience and artistic process. For instance, see how an artist from the Barcelona collective speaks of her work:

> My work focused on engravings in the beginning. <…> And the process of engraving has a lot of test samples that are left all over the workplace; and the process started to interest me a lot. So, I made a work where I tried to look at nine samples that I was

> working on as the final result. That is when I found this enthusiasm. (BF[2], artist, interview)

An artist from the Madrid collective, although referring to joint creative practice typical for the collective, reproduces the same position-specific orientation, also speaking of individuality and creative process:

> The public space allows for interaction between individuals through their own singularity. We can make part of a community using our individuality, summing up synergies to launch common projects that would diversify virtual collective space. The web allows us to enjoy heterogeneous spaces of open access and make part of their conception and creation. This philosophy, centered in collective participation, open access and freedom of use has given place to a new type of society, more solidary and generous, where all kind of professionals share their knowledge and make it work for free the common good. (MB, artist, newspaper interview)

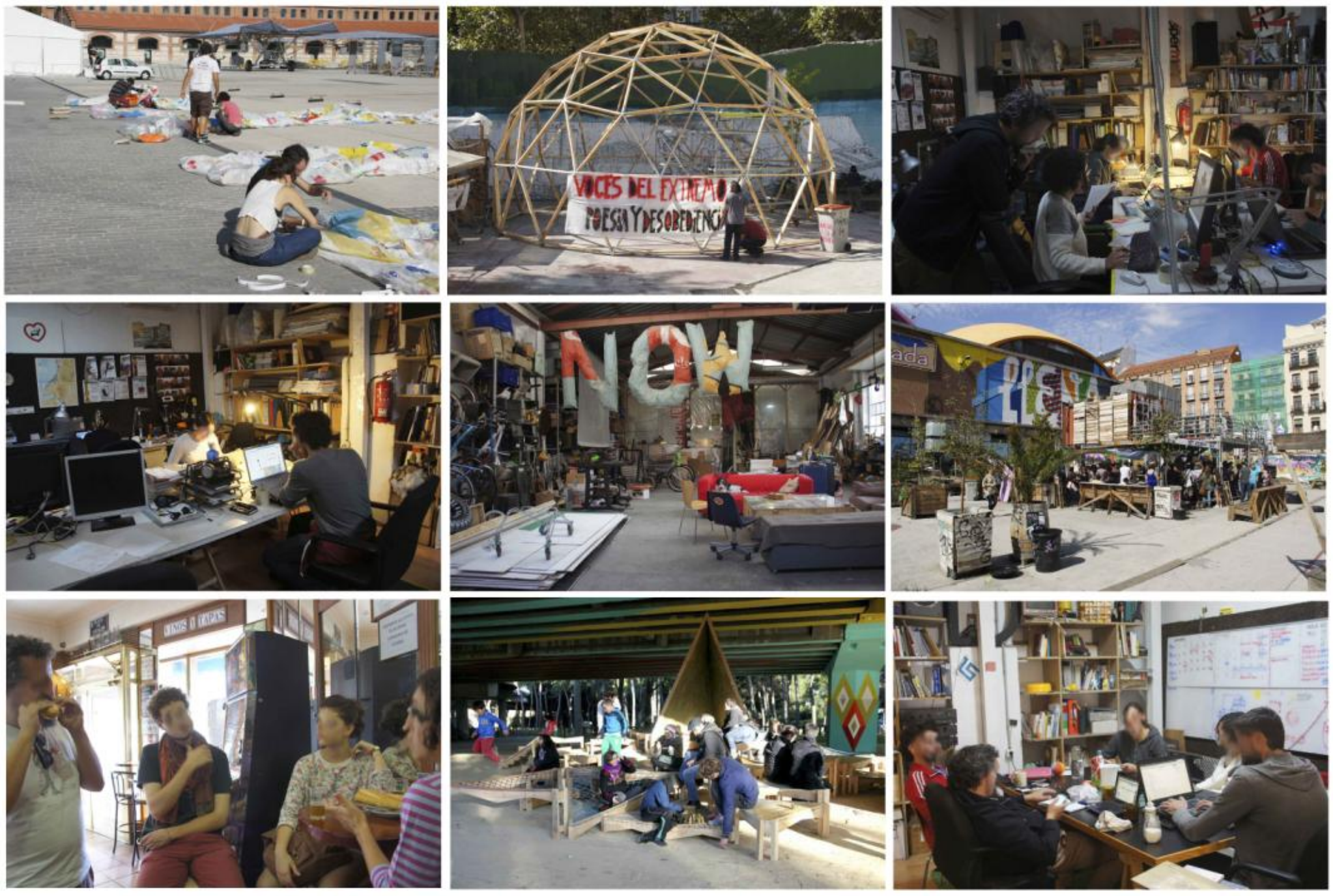

**Figure 2**. The Madrid collective.

[2] Here and further: The first letter in an acronym for a member refers to the collective, 'M' standing for Madrid and 'B'—for Barcelona, the second letter being a randomly assigned code.

Managers, on the contrary, aim at market success for their collectives, which they link to visibility, good funding, formal partnerships, careful budgeting and timing, and occupation of an empty market niche:

> The problem is to find a new form of communication with enterprises, have a conversation with them. <…> We need to make it more efficient and make this relationship the best it can be. Or we continue on our own and stay in the middle of nowhere. We are small, not a large NGO or a foundation or whatever. We need partners all the time. (MB, manager, interview).

> The evaluation meetings that we have from time to time as well as other meetings, like the one after this summer, are to see how we are doing, to revise our budget every three months. <...> To see if we need to push things forward to get more projects or we can relax and make some investments into the future. <…> I think it is appreciated. Evaluation. This is not like a creative session, for example, not all of us are here, like, for example, in a project where we have to design a seminar in public spaces for European civil servants. (MG, manager, interview)

> Starting [collective's name] required us to make contact with centers of reference within the world of art and education, see how the dynamics of the system function and get to know different visions around educational process. Along all this trajectory of construction we have seen unions between museums of art and schools as well as artists and schools, but we have also detected an absence that we believe might be a shared desire: the union school—school / classroom—classroom. (BI, manager, interview)

In accordance with their orientations, 'artists' and 'managers' opt for different functions in the collectives. The former focus mainly on their creative work while the latter coordinate projects, attract funding for the projects from companies, states, and regions, do promotion, accounting and operational planning, administer the space, coordinate cooperation with partners in the artistic scene, and work with broad nonprofessional audiences. The members also express discontent to the duties inconsistent with their field positions. For instance, in Barcelona, while encouraged to participate in the educational activities, autonomy-oriented members unanimously prefer to work on their solo artistic projects. Indeed, educational work does not promote the career of an autonomous artist. In addition, the members tend to be friends and collaborate mainly with those occupying similar positions. 'Artists' jointly exhibit, coordinate their creative processes, use the same objects and techniques to inspire their creativity, and frequently hang out together. 'Managers' team up to work with broad audiences, organize workshops, teach, run the shared space, cooperate with artistic and educational institutions, do accounting, write reports, and apply for funding. They also tend to casually interact mainly within their circle.

Simultaneously, both collectives are highly informal small groups, in which members share spaces of work and leisure, frequently interact, engage in joint creative and everyday activities, and become friends not only within but also across field positions. The Barcelona collective conducts regular informal reunion meetings of all members, organizes joint work on public events such as open door days, engages the artists in educational activities run by the managers, and so on. In the Madrid collective artists and managers constantly work jointly to develop their complex collective projects that involve architectural production, public space interventions, and accompanying events. Hence, in each of the collectives we can also expect local group cultures to emerge across the positions.

## 5. Dataset

The data for this analysis were gathered in 2014, through two parallel in-depth ethnographic studies conducted in each of the collectives following uniform procedures. The dataset includes triangulated social network survey data and multisource textual data, such as interviews, expressions in dialogues, and texts written by the members.

Only core members with enduring membership and deep engagement in group practice were included in the analysis. All of these members agreed to participate in the study.

To gather social network data, the members were asked to report on two types of their social network ties—collaborations and friendships—by responding to a roster-based social network questionnaire. To report on collaborations, individuals marked those of the other members, who they frequently worked with during six months prior to the study. This usually involved joint creation of artworks, preparation of exhibitions, project coordination, administrative work, or fundraising.  Friendships were reported by marking those who they felt strongly emotionally attached to. Then, the answers were compared to find the ties confirmed by both in a dyad, and the networks were plotted so that ties indicated by only one person in a dyad were displayed as dotted lines. Presenting the plots to each of the other members, field researchers asked them if the dotted lines actually exist. Only the ties confirmed by the majority of the core members were kept in the resulting networks.

The main body of the textual data are transcripts of interviews with the members. The interviewers asked each member up to 50 open-ended questions from a uniform guide, such as: “What is art?”; “What are you currently working on?”; “How do you see the contemporary artistic scene?”, necessarily covering a uniform set of core topics. The interviews concluded by asking about members' statuses, occupations, affiliations, biographical and career trajectories, and education. The 20 interviews lasted from 30 minutes to 2 hours.

To enrich the textual corpora, field researchers collected texts written by the members, including texts on the collectives' webpages, annotations to artworks, and papers in media. The texts cover various topics, from reflections on creativity and descriptions of creative processes and practices to comments on the state of the artistic scene and the social, cultural, political, and economic context of their collective.

In addition, the textual corpora includes transcripts of members' narratives collected during (a) unstructured parts of the 20 photoelicitation procedures conducted with each member, when they commented on the photos of shared spaces and objects, and (b) 17 visual ethnographies conducted in the artistic studios and other spaces where joint practice of the collectives extended.

Most of the textual data were collected in Spanish. The texts in Catalan were translated into Spanish prior to inclusion in the dataset.

The sizes of the 20 resulting individual text corpora vary between 7,156 and 20,916 words. For some members there is more textual material than for others, which corresponds to the fact that some individuals are more vocal in expressing their cultural orientations and, hence, have a greater chance to develop cultural similarity to others.

To distinguish between the autonomy-oriented 'artists' and the heteronomy-oriented 'managers', the information signaling members' objective and subjective positions in the field of cultural production from the interviews was combined with the other ethnographic data on practices, interactions, and extra-collective relations of the members. Additional background information was collected from other members and online. These data were analyzed by the field researchers who have in-depth knowledge and intuitive understanding of the two empirical settings.

# 6. Cultural Similarities Within and Between Positions

This section starts the analysis by comparing cultural similarities between 'artists' and 'managers'. I compare both within and across the positions and the collectives. To account not only for the field-imposed culture, but also for the cultural similarities emerging from the context of collective practice, I must avoid the approaches that capture position-affected culture, whether *a priori*, as close-ended survey questions do (e.g., Yeung, 2005; Lizardo, 2006; Vaisey and Lizardo, 2010), or *a posteriori*, as coding implies (e.g., McLean, 1998). Instead, I apply a semantic network analysis machine text processing technique that traces associations between concepts (word stems) based on the collocation of words in texts (Danowski, 1982; Carley, 1986, 1991; Nerghes, Hellsten and Groenewegen, 2015; Antonyuk, 2018). *Mapping* words and associations the ways they are expressed by individuals along the local context of their practice (for a discussion of mapping v. coding see Lee and Martin, 2015) enables capturing both culture imposed to field positions and emergent culture that refracts the imposed orientations (Basov et al., 2018).

Taking individual corpora of oral and written expressions by each of the 20 members of the two collectives, Automap (Carley, Columbus and Azoulay, 2012) was applied to produce 20 semantic networks including the concepts a member used and the associations between them she expressed. After the removal of stop words, such as articles, particles, prepositions, and pronouns, Porter's stemming procedure collapses variants of the same word, e.g., singular and plural forms, to their stems by removing suffixes (Porter, 1980) to yield concepts—the nodes in semantic networks. To obtain an individual semantic network, an undirected link was created between every two concepts which appear immediately next to each other in a sentence (i.e., are not separated by a full stop mark or by another word) in the individual's expressions.

The 20 individual semantic networks resulting from these procedures ranged from 743 concepts connected with 225 associations to 1,697 concepts and 1,203 associations. Out of these, only stable associations, i.e., those encountered at least twice in an individual's verbal expressions, were selected (for a more restrictive approach to selection of semantic cores, see for instance: Pavan and Mainardi, 2018). This yielded individual networks sized from 25 concepts connected by 13 associations to 141 concepts and 112 associations.

Comparing expressed concepts and associations, one can conclude on cultural similarities between the individuals. For instance, Figure 3 shows associations that two members of the Madrid collective, ME and MH, have with the concept *art* shared by these members. The plot not only shows that members mostly differ in the concepts they use, but also that they differ in their perspectives on art: ME views art as 'contemporary' and 'pure', while MH associates art with 'public'. These differences signal the orientation of ME at cultural autonomy and of MH—at the heteronomy.

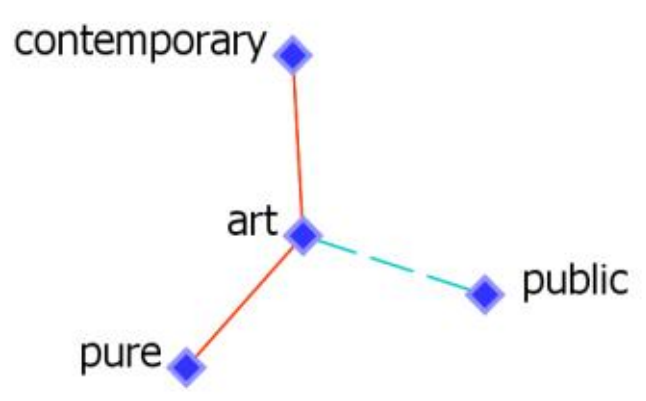

**Figure 3.** Associations of the concept ‘art’ shared by ME and MH in the Madrid collective.

*Note: Diamonds=concepts; longdashed line=association of MH; solid lines=associations of ME.*

To check how culturally similar members of the collectives are in the same and in different field positions within and across the collectives, the following procedure was applied. Firstly, Jaccard indices are calculated to evaluate the pairwise similarity of members’ semantic networks across the two collectives (see also Basov and Brennecke, 2017). This was done in two alternative ways: by concepts and by concept associations. Whereas the former evaluate the similarity of members’ vocabularies, the latter capture similarities in the ways members understand concepts. Then, the two types of dyadic similarity coefficients were put in two separate square matrices and multidimensional scaling (MDS), a technique conventional in formal analysis of culture (McLean, 1998; Ruef, 1999; Christopoulos and Vogl, 2015), was applied to plot each of these matrices so that geometric distances between nodes indicate cultural similarity between members. The results are presented in Figure 4.

MDS shows that whether semantic similarity was treated as usage of similar concepts or as usage of similar concept associations, members occupying different positions within collectives are more culturally similar than members of different collectives occupying similar positions. This outcome meets the assumption that cultural similarities imposed by the field are, quite effectively, contested by cultural similarities emerging in groups.

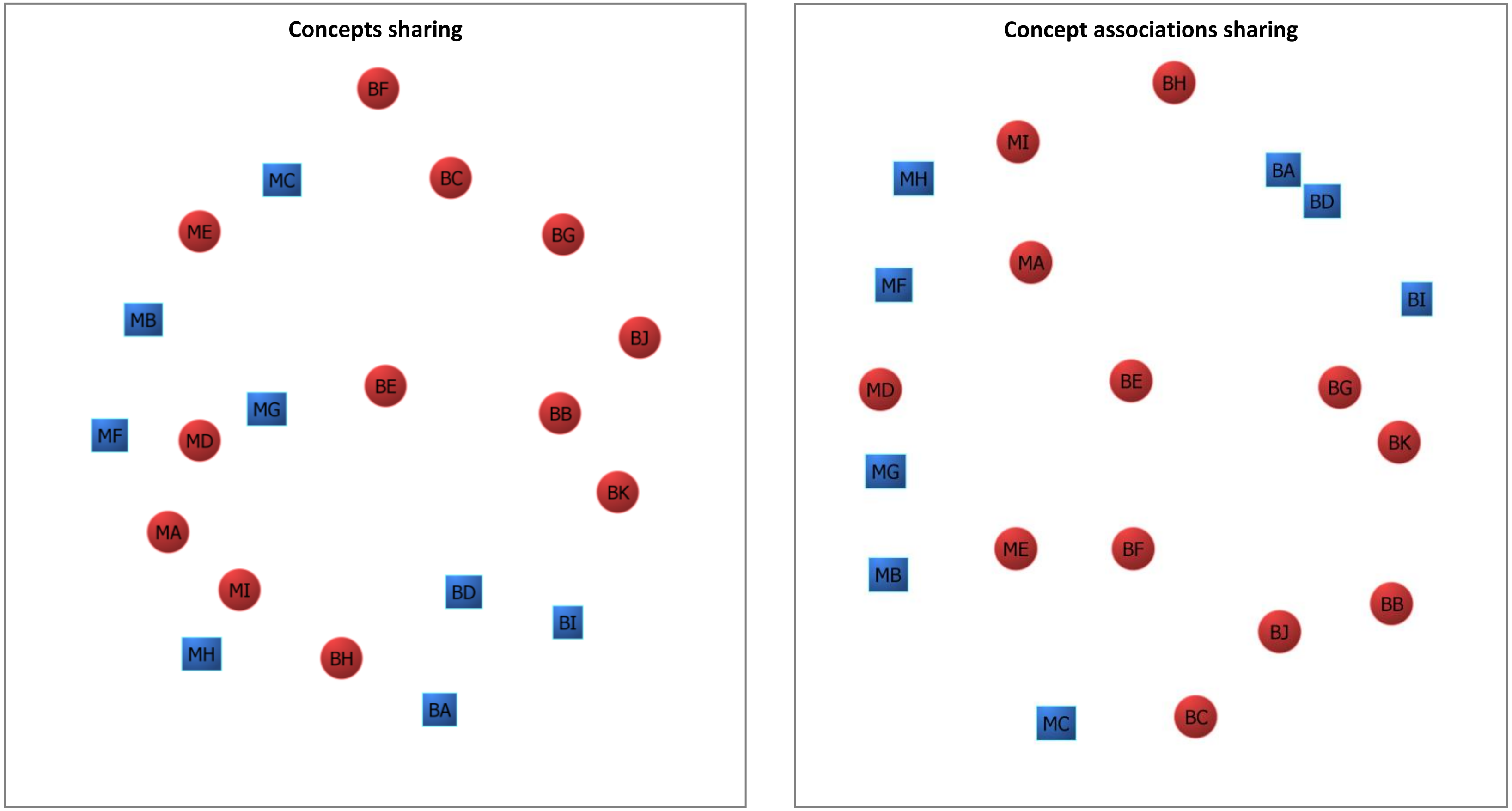

**Figure 4.** Semantic similarities between members of the collectives.

*Note: Left—similarities evaluated as concept sharing; right—similarities evaluated as sharing of concept associations; squares=managers; circles=artists; 'M' in the node title indicates Madrid members and 'B'—Barcelona members.*

# 7. Cultural Homophily Within and Between Field Positions

To examine cultural homophily, collective semantic networks for each of the two cases were produced by unioning the individual semantic networks of their members (comp., e.g., with Pavan, 2012: 121), and combined with social networks of friendship and collaboration based on the 'full' version (Lee and Martin, 2018) of the two-mode approach (Breiger, 1974; Fararo, 1984; Malinick, Tindall and Diani, 2013) drawing on concept usage by individuals[3]. This resulted in two-layer socio-semantic networks of each collective, including:

- a *semantic network* of concept associations, as concepts are collocated in the expressions of members;
- a *social network* of friendships and collaborations;
- a *concept usage* network connecting the social network and the semantic network.

Figure 5 plots the two resulting socio-semantic networks. The key descriptive statistics on the networks, including individual attributes, are summarized in the Online supplementary materials, Appendix A.

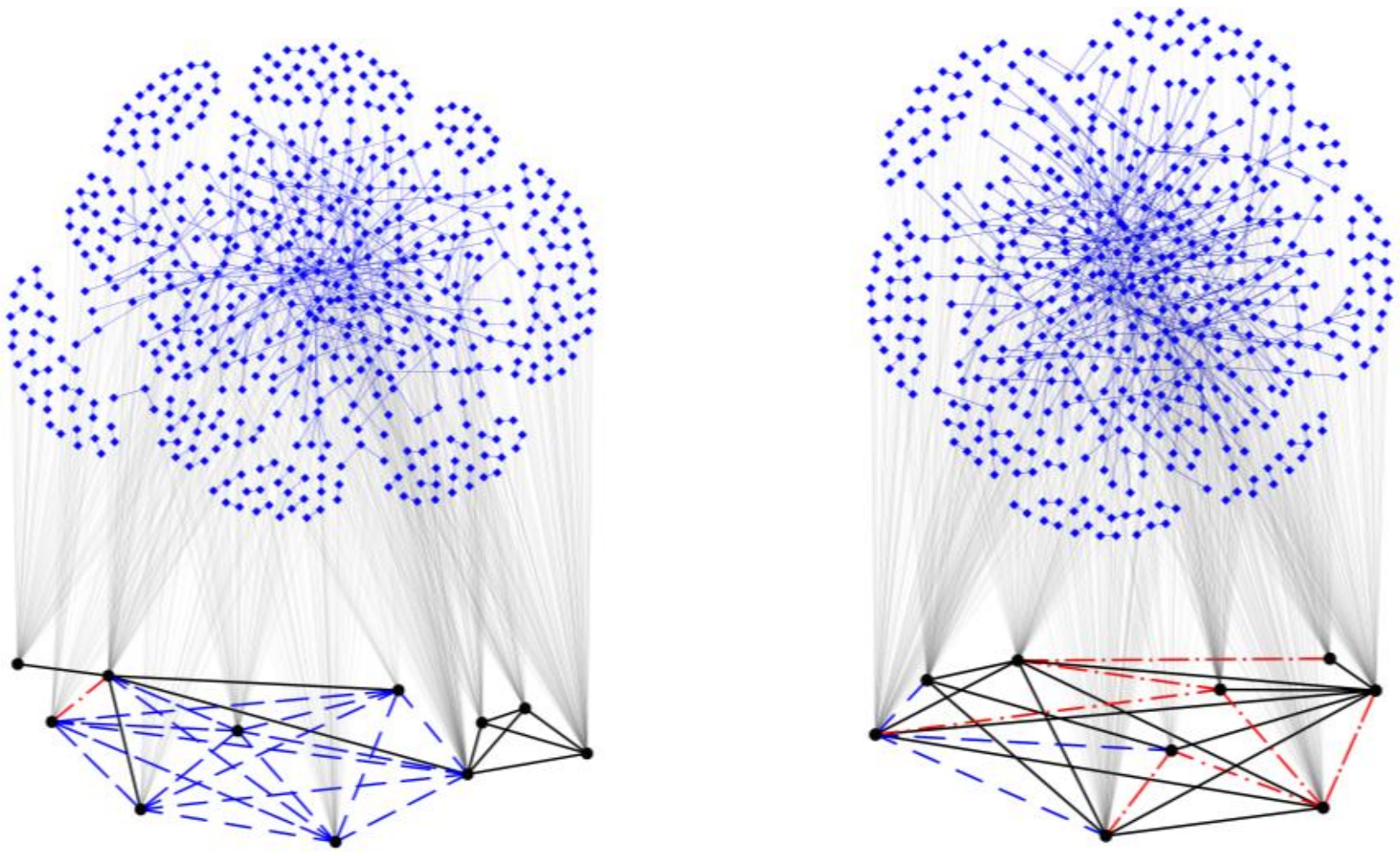

**Figure 5.** Socio-semantic networks of the Barcelona (left) and the Madrid (right) collectives.

*Note: Diamonds=concepts; circles=members; lines between diamonds and circles=concept usage; lines between diamonds=concept associations; solid lines between circles=multiplex social ties combining friendships and collaborations; longdashed lines=collaboration ties only, dotdashed lines=friendships ties only*.

[3] This is only one of the possible 2-mode approaches to linking nodes in the cultural structure to social networks. Alternative approaches were proposed, for instance, by Roth and Cointet (2010), Brennecke and Rank (2017), Basov, Lee, and Antoniuk (2017), and Godart (2018).

Now, to test for cultural homophily, specific patterns in the socio-semantic networks can be analyzed, such as (1) socio-semantic triangles that include a social network tie and a concept two individuals share, and (2) socio-semantic cycles including a social network tie and a concept association two individuals are affiliated with. For instance, when two friends in the Barcelona collective, BD and BA, both use the concept *politics*, we have a socio-semantic triangle, and when they use associated concepts *education* and *artistic*, this yields a cycle (see Figure 6). Examining ensembles of such patterns in socio-semantic networks while fixing semantic and concept usage networks, hence treating culture as an independent variable, we can infer cultural homophily, i.e., how cultural similarities, operationalized as similarities in concepts and concept associations individuals express, affect the observed social networks.

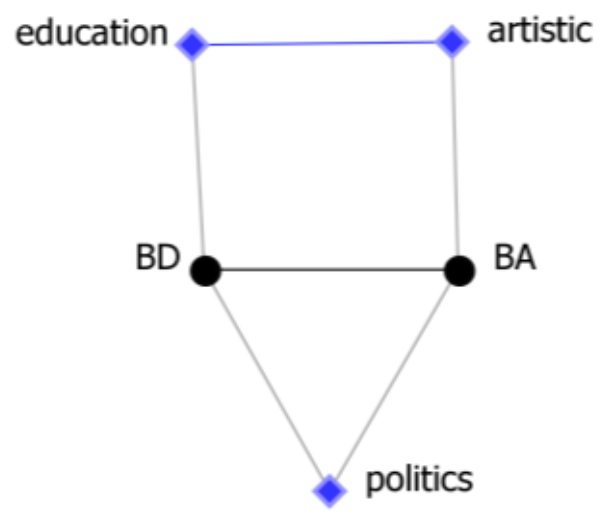

**Figure 6.** A socio-semantic triangle (bottom) and a socio-semantic cycle (top) in the Barcelona collective.

*Note: Diamonds=concepts; circles=members; lines between diamonds and circles=concept usage; line between diamonds=concept association; line between circles=social network tie.*

Furthermore, formalizing positions of individuals in the field of cultural production as actor attributes (for similar approaches, see Lizardo, 2006; Godart and Mears, 2009), I can take field positions into account during my assessment of whether cultural homophily reproduces field structure or not. Note this approach to maintain the distinction between objective relations and social network ties, as advised by Bourdieu.

The tests for the cultural homophily mechanism under the impact of field while taking into account other social network mechanisms as well as the mechanisms operative in the object usage networks, are possible thanks to the recently developed multilevel exponential random graph models (Wang et al., 2013). ERGMs in general aim at capturing the probability to observe the empirical network as a set of parameter estimates associated with different network patterns corresponding to certain mechanisms of network evolution. For instance, triangles correspond to the triadic closure mechanism and, estimating parameter value for triangles in the context of other patterns, ERGMs capture the relative importance of triadic closure in comparison to other mechanisms in yielding the observed network. MERGMs are a new class of ERGMs addressing network links not only within, but also between and across dually related networks of different types by modeling two one-mode networks connected with a third, two-mode, network. Successfully applied to model two-level networks of social ties and knowledge elements (Brennecke and Rank, 2017), these models fit the purpose of studying two-layer socio-semantic networks, where social ties are linked to both semantic concepts and concept associations—via usage links between individuals and concepts. Simultaneously, like any ERGMs but unlike regression-based modeling approaches, MERGMs assume that the presence or absence of specific links affect the presence or absence of other links. This allows a structural account of complete social and semantic networks when examining the impact of culture on social ties (compare with Vaisey and Lizardo, 2010).

I use MPNet (Wang, Robins, Pattison and Koskinen, 2014), a software tool designed to conduct MERG modeling via Markov chain Monte Carlo maximum likelihood estimations (Snijders, 2002). Because my aim is to test how dyadic social ties between members of the collectives in similar and different field positions are affected by concepts and concept associations both of them are affiliated with, I technically treat collectives' cultural structures as exogenous determinants of their social networks. This is done by 'fixing' semantic and concept usage networks in the model estimations (see also Basov, 2018). This function is featured by MPNet to "treat one or more of the three networks involved in the two-level network as fixed and exogenous" (Wang et al., 2014: 20). I produce separate models for the two types of social ties: friendships and collaborations.  I analyze the networks observed in the two collectives in an aggregated fashion using 'structural zeros' technique, which disavows nominations of ties between the collectives during modeling (Kalish and Luria, 2013; Matous and Wang, 2019).

Results of modeling are presented in Table 1. There, the focal, *Full socio-cultural*, models contain the main effects of interest intended to unveil how cultural homophily works within and between field positions. To check for the effect of concept sharing on ties between individuals in similar positions and in different positions, I account for socio-semantic triangles with actor attributes. To test for the effect of concept associations sharing on ties between individuals in similar and in different positions,

I include socio-semantic cycles with actor attributes. Additionally, I account for several other network structuring mechanisms as controls that might have a parallel effect on social ties formation, including degree distribution and triadic closure in social networks, the immediate effect of field position similarity on tie formation, and cultural homophily irrespective of field positions.

To ensure that cultural homophily and field positions matter, additionally to the *Full socio-cultural* models I estimate separate models for: (a) exclusively social networks (*Basic social*), (b) social networks with field effects (*Full social*), and (c) social networks with cultural and positional effects (*Socio-cultural*).

Among positive and significant effects in the *Full socio-cultural* models are those for the patterns with shared concepts and concept associations. The modeling results, thus, suggest that social network ties, both friendships and collaborations, are affected by cultural similarities, in accordance with the cultural homophily mechanism. Note that only those of the effects of cultural similarities on social ties that account for field positions are significant in these models. Although I find some influence of culture on social network ties in the *Socio-cultural* model for friendship (*Influence of concepts usage on establishment of social ties by individuals*: 0.030(0.015), $p<.05$), it is not any longer so in the *Full socio-cultural* model for friendship.

More specifically, in both Friendship and Collaboration *Full socio-cultural* models, parameters for the effect *Influence of concept sharing on establishment of dyadic social ties within field positions* are positive and significant (friendships: 0.237(0.110), $p<.05$; collaborations: 0.288(0.100), $p<.05$), whereas parameters for the impact of concepts sharing on dyadic ties *between* field positions are not significant. It indicates that sharing of single concepts (having similar vocabularies) stimulates collaboration and friendship ties between individuals in similar field positions, but not in different positions. This result suggests that cultural homophily reproduces objective relations in the field of cultural production at the level of intersubjective social network ties.

In contrast, parameter for the effect *Influence of joint engagement with concept associations on establishment of dyadic social ties within field positions* in the Full socio-cultural models is not significant, whereas *Influence of joint engagement with concept associations on establishment of dyadic social ties between field positions* is positive and significant for collaborations (0.114(0.051), $p<.05$). It indicates that affiliations with the same concept associations stimulate collaborations between artists and managers and thus, drawing on the concept associations, cultural homophily contests field forces.

Table 1. Results of MERGMs.

| | | | Friendship models | | | | | | | | Collaboration models | | | | | | | |
|---|---|---|---|---|---|---|---|---|---|---|---|---|---|---|---|---|---|---|
| *Type(s) of independent variable(s)* | *Effect* | *Illustration* | *Full socio-cultural* | | *Socio-cultural* | | *Full social* | | *Basic social* | | *Full socio-cultural* | | *Socio-cultural* | | *Full social* | | *Basic social* | |
| | Baseline tendency to establish social ties | 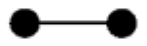 | -5.826 | 8.205 | -12.600 | 12.307 | -1.644 | 0.988 | -1.821 | 0.968 | -4.409 | 2.670 | -1.851 | 2.354 | -0.882 | 1.594 | -0.383 | 1.494 |
| | Degree distribution in social network | 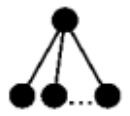 | -0.571 | 0.923 | -0.870 | 0.798 | -0.207 | 0.505 | -0.244 | 0.504 | 0.064 | 0.783 | -0.951 | 0.649 | -1.047 | 0.575 | **-1.215*** | 0.591 |
| | Tendency to form social triads | 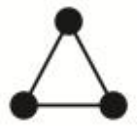 | 0.624 | 0.524 | 0.699 | 0.472 | **0.897*** | 0.428 | **0.956*** | 0.453 | **1.818*** | 0.556 | **1.675*** | 0.556 | **1.626*** | 0.529 | **1.868*** | 0.550 |
| | Tendency to establish social ties with those in similar positions | 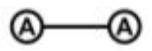 | -0.016 | 0.572 | 0.148 | 0.562 | -0.326 | 0.469 | - | - | **0.982*** | 0.427 | **1.026*** | 0.366 | **0.972*** | 0.344 | - | - |
| Cultural | Influence of concept usage on establishment of social ties by individuals | 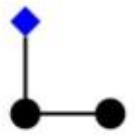 | -0.012 | 0.059 | **0.030*** | 0.015 | - | - | - | - | -0.089 | 0.05 | 0.003 | 0.009 | - | - | - | - |
| | Influence of concept sharing on establishment of dyadic social ties | 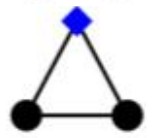 | 1.958 | 4.492 | 5.518 | 6.502 | - | - | - | - | -0.082 | 0.792 | -0.105 | 0.716 | - | - | - | - |
| | Influence of joint engagement with concept associations on establishment of dyadic social ties | 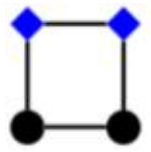 | 0.010 | 0.056 | -0.023 | 0.017 | - | - | - | - | 0.091 | 0.053 | 0.002 | 0.013 | - | - | - | - |
| Cultural and positional | Influence of concept sharing on establishment of dyadic social ties within field positions | 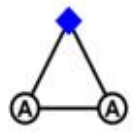 | **0.237*** | 0.110 | - | - | - | - | - | - | **0.288*** | 0.100 | - | - | - | - | - | - |
| | Influence of concept sharing on establishment of dyadic social ties between field positions | 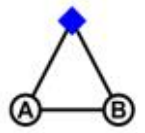 | 0.112 | 0.094 | - | - | - | - | - | - | 0.039 | 0.076 | - | - | - | - | - | - |
| | Influence of joint engagement with concept associations on establishment of dyadic social ties within filed positions | 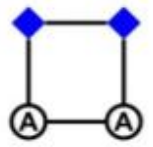 | -0.018 | 0.055 | - | - | - | - | - | - | 0.005 | 0.045 | - | - | - | - | - | - |
| | Influence of joint engagement with concept associations on establishment of dyadic social ties between field positions | 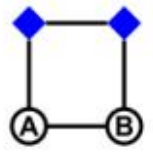 | 0.040 | 0.063 | - | - | - | - | - | - | **0.114*** | 0.051 | - | - | - | - | - | - |

*Note: Unstandardized coefficients; two-tailed tests reported; *p < .05. For the presented models, I test the goodness of fit (GOF) with MPNet (Wang et al., 2014). The results are provided in Online supplementary materials, Appendix B. To ensure that the high Edge parameter in the Socio-cultural model for friendship does not affect the results, I re-estimate the model with the same set of effects, but fixed density. This yields the same outcome with respect to significance and GOF test still signals a good fit.*

*Diamonds=concepts; circles=individuals; lines between diamonds and circles=concept usage; lines between diamonds=concept associations; lines between circles=social network ties.*

To sum up, the cultural homophily mechanism is ambivalent. On the one hand, it reproduces the field by stimulating friendships and collaborations between similarly positioned group members who share similar vocabularies. On the other hand, cultural homophily contests field forces by stimulating collaborations between differently positioned group members who are affiliated with different concept associations.

Note the additional models without culture- and position-related effects to exhibit also the mechanisms well-known in social network analysis, such as triadic closure and centralization (see parameters for *Degree distribution in social network* and *Tendency to form social triads* in *Basic social* and *Full social* models). However, most of these effects become insignificant as I start accounting for the impact of cultural similarities moving into the *Full socio-cultural* models.[4] This finding suggests that social network modeling may need to account for culture in order to arrive at accurate results, as many of the processes traditionally in the focus of social network analysis may result from cultural effects. In this, my results complement the recent arguments and findings on the impact of cultural worldviews and tastes on social networks (Lizardo, 2006; Vaisey and Lizardo, 2010), expanding these findings to cultural structures. This is not all, however. Although, in line with these studies, my *Socio-cultural* model for friendships does, indeed, show an effect of engagement with group culture on ego's friendship ties, *Full socio-cultural* models for both friendship and collaboration show no effect of cultural constructs on social ties unless field positions are accounted for. Hence, the recent findings on the effects of cultural worldviews and tastes on 'egocentric' social networks might benefit from additional testing based on 'sociocentric' data (for a similar argument see Lewis and Kaufman, 2018) while accounting for field positions. An alternative explanation is that, unlike concepts and associations freely expressed by group members in the present study, closed-ended survey questions on cultural worldviews and tastes already impose field positions.

[4] Compare, for instance, with Block (2015), who finds the tendency against three-cycles to be spurious if one controls for reciprocity in friendship networks. Similarly, previous results of socio-material MERGMs on art collectives, confirm this for friendships and, in addition, show the same for collaborations (Basov, 2018). In the current models, however, while it still holds the same in the 'social' models, when I include the effect of culture in the 'full' and even in the 'socio-cultural' models, it appears that triangles remain significant only in the collaboration networks but not in the friendship networks. This being said, Block (2015) is using SAOMs, not ERGMs, and the data is directed, while mine is not, so direct comparisons between the findings might be problematic.

## 8. What Is This All About? Examining the Cultural Content

The statistical analysis presented above is based on joint affiliations of individuals with concepts and associations in semantic networks of the collectives, but does not take into account the content of these concepts and associations. How do we know if they express position-specific culture or emergent culture? In this section, I take a closer look at the concepts and concept associations part of socio-semantic triangles and cycles found to indicate cultural homophily within and between field positions, correspondingly. Owing to space limitations, I present results on the Barcelona collective, but the findings on the Madrid collective are similar.

To begin with, I know from the models that sharing of concepts stimulates friendships and collaborations between members who occupy similar field positions. I examine these concepts and expressions in which such dyadically connected members use them—to see if they do signal the position-specific cultural orientations. Then, perhaps, I can conclude that homophily draws on position-specific cultural similarities to reproduce field structure in the social network of the collective.

In the socio-semantic network of the Barcelona collective, there are 80 unique socio-semantic triangles that include a shared concept and two similarly positioned friends or collaborators, either artists or managers (Figure 7). As the 80 shared concepts are too many to consider in detail here, I focus on ten used most frequently (Table 2), from the most frequent *work*, used 472 times in total, to *illustrate*, used 65 times in total. Among them, the highest frequency have the concepts part of socio-semantic triangles both of artists and of managers. These are the concepts that do not suggest any position-specific cultural orientations, but rather refer to the notions central for the whole field of cultural production, such as *work*, *artist(ic)*, *project*, and *art*. However, let us have a look at the quotes by the dyads using them. Several hundreds of quotes include these common concepts and here I provide only a few typical quotes as the examples, but many others reveal the same.

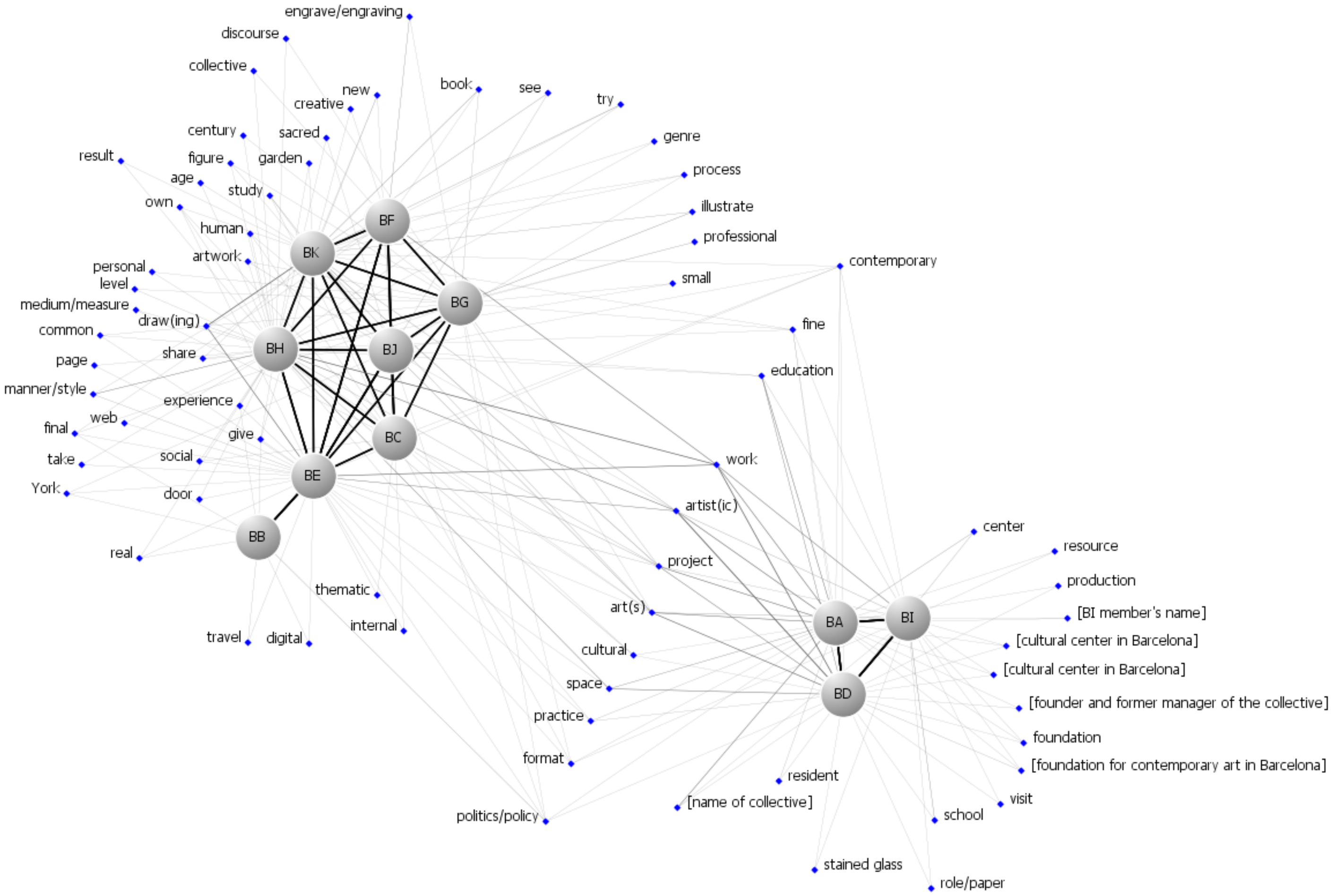

**Figure 7.** Socio-semantic triangles within positions in the Barcelona collective.

*Note: Diamonds=concepts; lines between diamonds and circles=concept usage; lines between circles= social ties (friendships or collaborations) within positions (ties between positions are omitted); BA, BI, BD—managers; BB, BC, BE, BF, BG, BH, BJ, BK—artists; line width corresponds to frequency of usage of a concept by a member. Concepts with equivalent translation were merged in the plot for the sake of readability.*

**Table 2.** 10 most frequent concepts shared by dyads of similarly positioned individuals.

| Concept | Times used | | |
|---|---|---|---|
| | *artists* | *managers* | *total* |
| work | 241 | 231 | 472 |
| artist(ic) | 156 | 220 | 376 |
| project | 142 | 150 | 292 |
| art | 84 | 147 | 231 |
| draw(ing) | 201 | 0 | 201 |
| space | 61 | 90 | 151 |
| education | 10 | 129 | 139 |
| politics/policy | 71 | 17 | 88 |
| manner/style | 83 | 0 | 83 |
| illustrate | 65 | 0 | 65 |

Consider the quotes by a dyad of artists with the most frequently used concept *work*:

> If you **work**[5] at home or in a workshop you **work** alone. According to an old French professor, artistic creative process is individual. It is one of inclusion. You are alone, you create and you only open to others when you need to show your **work**, when you communicate what you want to express. I was, in contrast, interested in the fact of making part of a network of a residence. Because of this coexistence when you are here, **working** with other artists, very different from you. Maybe you don't share ideas or styles of **work** but the fact of sharing space with other artists who have their own process of **work** [is important]. (AH[6], interview)

> When you **work** in art, you are in your studio and it is very lonely **work**. You are always in your own world. (AJ, interview)

Meanwhile, dyads of managers are far from viewing work as a process of individual artistic creation. Instead, they express orientations towards teamwork aimed at promotion in the market and community engagement:

> And the change for the better is not done by myself alone. The team is of a fundamental importance here, because oftentimes it is difficult, because one needs to learn how to make dossiers, how to present them, how to define ourselves, how the sector sees us. It has been hard **work**, well, it is hard **work**. (MD, interview)

[5] Here and further: Emphasis mine.
[6] Here and further: The first letter in an acronym for a member refers to the position: 'A' for autonomous and 'H'—for heteronomous and the second is a randomly assigned code.

> [On her educational work with children] In principle, I want to make them get together, try shared **work**, be a group. But to be individuals is also fine. I actually have problems with the word ‘creativity’. For me the objective is dialogue, listening to each other, respect. (MI, interview)

The subsequent quotes with the concept *artist* by two managers, who are both friends and collaborators, are quite vivid in expressing the perspective of managers on artists:

> How we work with an **artist**: We can work with an **artist** who explains their work or does not, but we can always try to explain, to question, to reread it. In many projects it is important that the **artist** is present. Because of the respect that we have for the creator and the processes of creation, we are interested in the pupils coming and seeing the everyday life of an **artist**, their space of creation. With the idea of proximity, of taking the **artist** down from the pedestal. (MD, interview)

> We have a catalogue with activities that we offer and the **artists** who participate and we offer it to schools. They can reserve a group to come and get to know the **artist**s, take part in an activity. If because of infrastructure it is better that the **artist** comes to the center, we can also do it. (MA, interview)

The quotes show that even the concepts common for triangles of artists and managers are used by them to express different cultural orientations specific to their positions. However, among the most frequently used concepts in Table 2, several are used only by dyads in one of the positions, but not the other (e.g., *draw(ing)*, *illustrate*, *manner/style*). Let us consider the most frequent of such position-specific focal concepts (Table 3).

These concepts directly name issues important for each of the positions. Whereas ties between autonomous artists are stimulated by shared references to novelty, experiment, and personal style (*new, try, manner/style*) and artistic products or activities (*engrave/engraving, draw(ing), book, and illustrate*), ties between managers are reinforced when both of them use the name the collective, names of foundations and cultural institutions the collective works with, name of a colleague manager who founded the collective, resident artists and visitors to the residence. Note also that artists tend to jointly refer to processes of doing something, while managers—to objects and names.

The quotes contextualizing these concepts confirm the impression. Consider expressions of two connected artists using the most frequent concept *draw(ing)* to define what this type of artwork is and to reflect on the corresponding creative process:

> This is a series of **drawings** where I tried to **draw** a plant… the process of **drawing** a plant. I did have a natural model of a plant, but it wasn’t the form nor the appearance of the

> plant that I wanted to capture on the paper. I focused on the moment of wanting to **draw** the plant. Because at the moment of **drawing** you have to focus on forms, on filled and empty spaces. You have to control and compare measurements to maintain the proportionality. So, the thing that I intended to **draw** here was precisely everything that I focused on to **draw** the plant and what doesn't appear in the **drawing**, what you can't see is the plant. I intended to **draw** the process of **drawing**. (AH, artwork elicitation)

> **Drawing** is one of the forms that assumes thinking in order to make itself present between us and our lives. It can be immediate as it allows you to take notes and sketch what you see and what you feel at a certain moment and place, as if it was a direct transposition of experience. It also has an intrinsic possibility of improvisation like performance and can be ephemeral and give a great sensation of creative freedom. In effect, the **drawing** is part of our relation with our physical environment when we register it and register in it the presence of a human. (AE, Master's thesis)

**Table 3.** 10 most frequent concepts shared by dyads in one of the positions, but not in the other.

| **Artists** | | **Managers** | |
|---|---|---|---|
| *Concept* | *Times used* | *Concept* | *Times used* |
| draw(ing) | 201 | [name of collective] | 97 |
| manner/style | 83 | [foundation for contemporary art in Barcelona] | 72 |
| illustrate | 65 | center | 68 |
| book | 64 | school | 51 |
| try | 55 | resident | 35 |
| see | 54 | visit | 26 |
| engrave/engraving | 52 | [founder and former manager of the collective] | 25 |
| study | 47 | [cultural center in Barcelona] | 22 |
| new | 46 | foundation | 20 |
| share | 45 | production | 19 |

Meanwhile, dyadically tied managers use the name of the collective to express their shared perspective on the collective as an art school, i.e., a cultural institution engaged in the popularization of art and sustaining relations with other institutions, something the autonomous artists in my dataset never reflect upon:

> **[Name of the collective]** is what we do. <...> Historically, there were various buildings and this one was **[name of the collective]** space. Sometimes there is a confusion. Now when we refer to **[name of the collective]** we refer to the association, not the building. So,

> there is the [artistic] residence, education, and the part of courses of seminars. (MA, interview)

> Starting **[name of the collective]** school obliged us to make contact with important centers within the world of art and education, see how the dynamics of the system function and get to know different visions around educational process. Along all this trajectory of construction we have seen unions between museums of art and schools as well as artists and schools, but we have also detected an absence that we believe might be a shared desire: the union school—school / classroom—classroom. (MI, interview)

Other quotes in the ethnographic data also confirm the assumption that single concepts are used by homophilous dyads in similar field positions to reproduce their position-specific orientations.

We also know from the models that artists tend to collaborate with managers when they jointly engage with concept associations. This binds the collective together despite the field-imposed distinction. Consider the plot in Figure 8 representing concept associations part of socio-semantic cycles of collaborators between positions. It is quite easy to see that most of such associations comprise an integral component, which perhaps, represents the core of group culture. Curiously, this core is built around the elements that we already know to be most frequently used in the collective overall to stimulate ties within positions: *artist(ic)*, *work,* and *project*. Thus, concept associations that stimulate social ties across positions connect concepts reinforcing social ties within positions.

Collaboration dyads of artists and managers in the Barcelona collective use 120 unique associations, in total. Five of them are used by all of the dyads between the positions: *artist(ic) work*, *artist(ic) project*, *artist(ic) education*, *education project*, and *fine art*. Used by members of the collective 1,968 times, altogether (Table 4), none of these associations, except for *education project,* refers to the topics specific only to one of the poles of the cultural production field (autonomous or heteronomous). We can, therefore, expect the quotes using these associations to express the group culture of the Barcelona collective that contests field-imposed culture and mediates between artists and managers.

**Table 4.** Concept associations present in all socio-semantic cycles of collaborators between positions in the Barcelona collective.

| **Concept association** | | **Times used** |
|---|---|---|
| artist(ic) | work | 607 |
| artist(ic) | project | 457 |
| artist(ic) | education | 418 |
| education | project | 418 |
| fine | art | 68 |

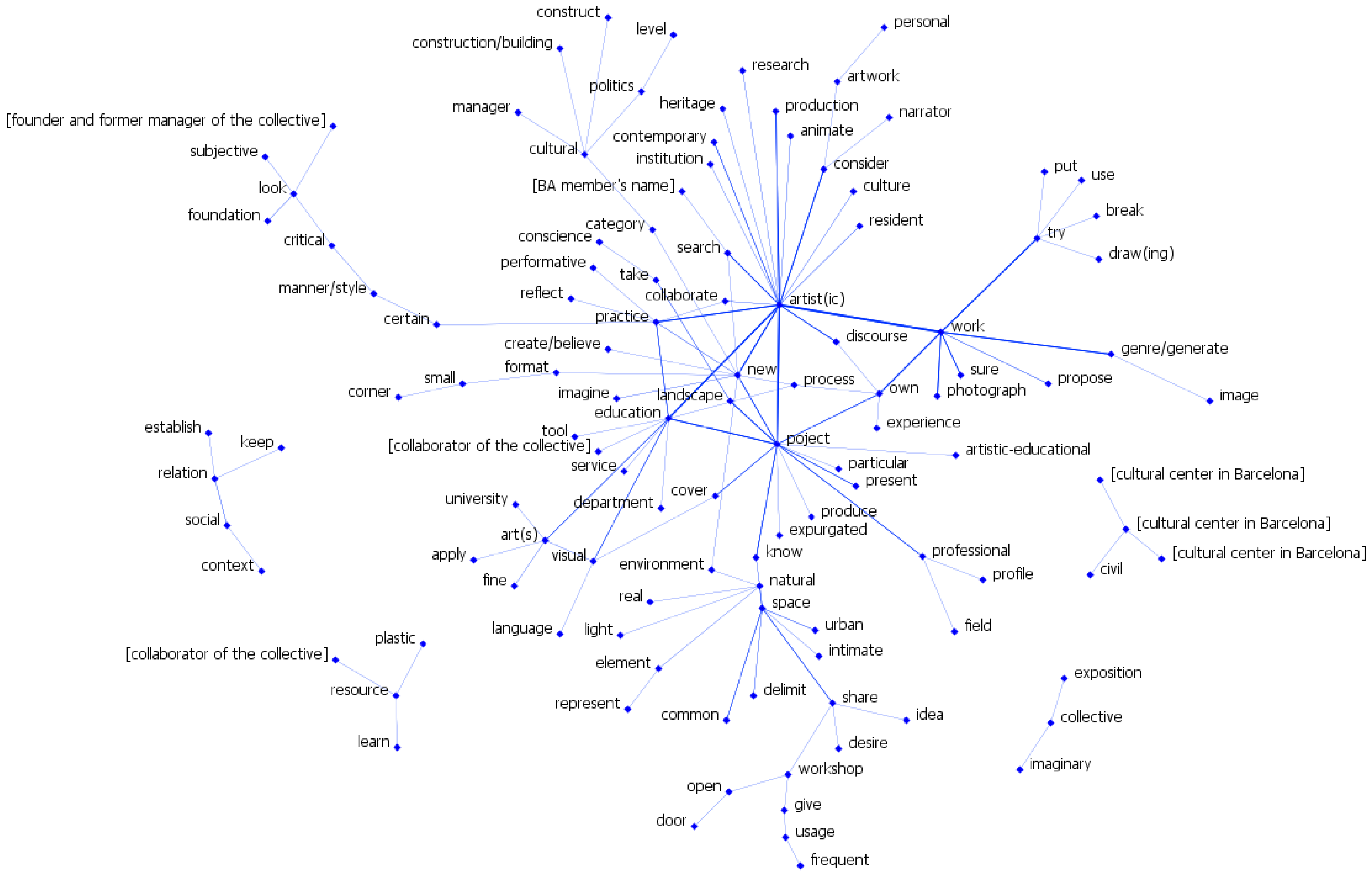

**Figure 8.** Concept associations part of socio-semantic cycles of collaborators between positions in the Barcelona collective.

*Note: Diamonds=concepts; lines=concept associations; line width corresponds to frequency of concept association in socio-semantic cycles between the positions. Concepts with equivalent translations were merged in the plot for the sake of readability.*

However, the ethnographic data do not support this assumption. Consider the following illustration of what *artistic project* is for collaborating artist AH and manager MD*:*

> [My project is] an **artistic project** articulated within the cultural concept of landscape and the approach to understanding of the surroundings, where one is situated, according to its origin and development in the history of art. A new approach comparing to the one I had at the beginning of my Master´s studies, which on the formal level takes shape in the liberty of strategies, as we can see in these two projects. The first consists in books that show my visual paths and the second—in compositions of image and text. Their meaning is an instinctive necessity to explore the context which surrounds me in order to discover, define and express my own experience in relation to natural as well as urban environments. It is not only about presenting these two projects, formulated and produced during my Master´s studies but also to create a memory of the evolution of the context of my artistic investigation. I expose the results of my work contrasting them to the approach and formal implementation of the beginning of Master´s, reflecting upon how my own work fits into the history of art that influenced me and the working methods of art characteristic of the last decades and seeing the possibility of continuity for the new orientation of my **artistic project**. (AH, project presentation)

> What I was doing was to update some **artistic projects** within this course [for Artistic Moderators or Animators] and to see how we can change it all together. I held the four sessions of the course on Saturday mornings and worked with institutions: museums, relations between schools and museums. Besides, I worked with artists who in social action or with **artistic projects** have to do with social relations. That was the topic that gave sense to the course. <...> Yes, to make artistic moderation, the course is called Course of Animators in **Artistic Projects**. (MD, interview)

And here is how they view *artistic work:*

> My **artistic work** does not take natural space of a landscape as canvas nor urban space as a scenario for action. I proceed from the pictorial tradition of landscape and experiment with the cultural category of landscape in order to make reference to the reality of a natural environment, systematically and without the necessary mimicry of topography. I take for my reference the processes that artists of Land Art and Matta Clark implemented to make a testimony of their processual experiences in natural or urban environment. Documents such as photographs of the place, maps, videos or movies, installations made with indigenous elements of the place where the intervention was carried out. In particular, photography to make a reference to the place. Similar to these artists and the long pictorial tradition of landscape, my **artistic work** emerges from the necessity to comprehend the place where I am and where I am developing socially. (AH, project presentation)

> "Open for Works" invites you to visit our space as changes are taking place. We are open so that you can see the works in progress before they are accomplished. To see the work of artists in their workshops and, most importantly, how and in what form they work. To amplify the preexisting ideas about the artist and their working processes. To facilitate the experimentation and critical observation in group or in family. To take you closer to contemporary art, showing the processes of production in a center of art education and promotion. To work with the stereotypes and typologies of artists to talk about them and

> present new models. To show **artistic work**, its tools, work space and results in relation to other professions and jobs. To relate the space of the workshop with a space of privacy. To present the workshop as a space of thinking and creation. To show the differences between the work done in the workshop by an artist and its public presentation. In different parts of the activity we will answer and develop the following questions: Who do you think is an artist? What materials do they use? What do you think their work space is like? What ideas and objects do they work with? How do they present them in public? During the visit the teacher offers different activities both in individual form and in a group. (MD, project presentation)

In a similar fashion, inspection of several hundreds of quotes delivers dozens of expressions by collaborating artists and managers that contain the focal shared concept associations, but do not blend orientations of different positions, instead communicating dissimilar position-specific orientations. In their narratives, dialogues, interview responses, and written texts, alike, artists speak of their individual expression, aesthetically bonded reflections, and art as a continuous creative process, whereas managers refer to goal-oriented joint work, organizing and practical issues, and describe artistic work as an end-product that is to be marketed. It makes a contrast to the results of statistical analysis, which shows concept associations to stimulate interpersonal ties between different field positions and not within positions.

# 9. Conclusion

Owing to the cultural homophily mechanism, cultural similarity stimulates social network ties between individuals. This implies occurrence or reinforcement of ties between individuals who occupy similar positions in social fields and hence are also culturally similar. Thus, cultural homophily is capable of reproducing field-level ‘objective relations’ at the intersubjective level of social ties. However, practice may contest field logic and bring together individuals occupying different field positions to generate group cultures. This may result in cultural similarities between individuals in different positions, leading to social ties between them through the very same mechanism of cultural homophily. My aim in this paper was to examine this ambivalent role of the cultural homophily mechanism by mixing qualitative, formal, and statistical socio-semantic network analyses in a study of two artistic collectives where different field positions meet and culture is both imposed and emerging.

I found intracollective cultural similarities across field positions to be higher than intercollective similarities within positions, which suggests that group culture emerging in practice contests culture imposed by the field. However, the cultural similarities within positions (at least, when it comes to similarities of vocabularies) appear to be sufficient for cultural homophily to reproduce the field in the groups. These stimulate friendship and collaboration ties between group members in similar field positions but not in distant ones. Nevertheless, I also found collaboration ties between different positions in the groups to be stimulated by being affiliated with the same concept associations. In other words, cultural homophily not only retranslates field forces to the micro level of social network ties, but also contests these forces; and it does so by drawing on cultural structures. This finding suggests that social networks may be not as obedient to culture and fields, as they have recently been argued by cultural sociologists (Lizardo, 2006; Vaisey and Lizardo, 2010; Bottero and Crossley, 2011). At least in small groups of joint practice, networks may also contest fields and the cultural orientations they impose (see also De Nooy, 2003; Basov et al., 2018), which is visible even when we focus on the effect of culture on social network ties, not the inverse effect.

These findings on the field of cultural production may very well apply to other fields revolving around creative activities, such as the academic field, and to groups that bring together different positions in order to pursue common practical goals, such as research teams. There, we can also expect cultural homophily to play an ambivalent role. Simultaneously, this might not work the same way for similar groups but operating in the fields more regulated by bureaucratic practice, such as small innovative firms in the economic field or collectives of political activists in the political field. Further research might compare the present findings with those on other fields using larger datasets—to see if my current results are generalizable.

The findings of this paper also suggest some methodological implications for socio-semantic network analysis in particular and, perhaps, for the studies of the relationship between the social and the cultural in general. Remember that, unlike the effect of shared vocabularies, the observed effect of concept associations is only visible when aggregations of these are statistically analyzed, but cannot be grasped intuitively from the members' verbal expressions, would these be interviews, written texts, or conversations. This is not owing exclusively to the well-known challenges of ethnographic data related to its volume and level of detail that hider focusing on the issues pivotal to a researcher (Sánchez-Jankowski, 2008; Abramson and Dohan, 2015): Even when we know where to look and examine the exact pieces containing the associations that stimulated social network ties, we find the informants to demonstrate conformity with their positions on the surface of their statements. Mediation of positions happens, literally, between the words, at the very basic level of concept associations reproduced across contexts. These associations are extremely difficult for a speaker/writer to consciously control and for a close listener/reader—to trace. The effect of such associations on social ties can be captured via formalization, aggregation, and application of statistical techniques, "superior to the human eye in pattern detection" (De Nooy, 2009: 48). Indeed, one should not expect the logic of practice to reveal itself explicitly in any single expression and capturing the effect of practice that mixes different universes of discourse requires totalization (Bourdieu, 1990: 82). Statistical socio-semantic analysis of formalized ethnographic data grants such a 'privilege of totalization' while remaining sensitive to practice. The analysis presented here is only one among many possible applications of the mixed socio-semantic approach, which has a lot more to contribute to our picture of the relationship between the social and the cultural.

**Acknowledgements**

The work presented here was supported by Russian Foundation for Basic Research (18-011-00796 “Dynamic of Sociocultural Network Structures“, 2018-2020) and The Centre for German and European Studies (Bielefeld University, St. Petersburg State University, and German Academic Exchange Service (DAAD) with funds from the German Foreign Office). The author expresses his gratitude to those who participated in collection and processing of the data used in this analysis: Aleksandra Nenko, Dafne Mutanyola, Artem Antonyuk, and Maria Drozdova. My special thanks are to Darya Kholodova, who helped in translating the quotes presented in the paper from Spanish and Catalan, and to Andrej Belyaev, who assisted in calculating similarity indices and in producing network plots presented in Figure 5. I am indebted to Wouter de Nooy and Ron Breiger for continuous theoretical and methodological discussions. Furthermore, I gratefully acknowledge the advice received from Peng Wang, Tom Snijders, and Garry Robins on applying MERGMs in this kind of analysis. I also cordially thank Aleksandra Nenko and Ksenia Puzireva for discussions of the findings and their friendly reviews on the earlier versions of the manuscript. In addition, this work has benefited from the comments of the participants of the seminar on “Socio-semantic patterns” in Paris, 2017. All errors and mistakes are my own.

# Appendices

## *Appendix 1. Descriptive statistics (average values across the collectives)*

| Statistic | Value |
|---|---|
| Concepts per group | 555.50 |
| Actors per group | 10.00 |
| *Semantic networks* | |
| Density | 0.29% |
| Average degree | 1.61% |
| *Concept usage* | |
| Density | 15.20% |
| Average concept degree | 1.49% |
| Average actor degree | 84.02% |
| *Friendship* | |
| Density | 39.20% |
| Average degree | 3.33% |
| *Collaboration* | |
| Density | 62.30% |
| Average degree | 4.34% |
| *Group member attributes* | |
| Female group members | 53.00% |
| Group members with artistic education | 84.35% |
| Variation in genre* | 57.85% |
| Cultural managers | 41.45% |

**Herfindahl index, 0=all members work in the same genre; 1= all work in different genres.*

## *Appendix 2. GOF statistics for MERGMs*

Conventional procedures (Hunter et al., 2008) included the test of effects in the models and a number of other known effects involving a calculation of t-values. Simulation conducted using the fitted models included a 100,000 iteration burn-in, followed by picking a sample of 1,000 networks from 100,000 iteration simulations. T-ratios for the effects not included in the models are close to 1.0 and for the effects parameterized by the models—to 0.1. These results indicate a good fit (Robins et al., 2009).

**Friendship models**

| | | *Full socio-cultural* | | | *Socio-cultural* | | | *Full social* | | | *Basic social* | | |
|---|---|---|---|---|---|---|---|---|---|---|---|---|---|
| Statistics | Observed | Mean | StdDev | t-ratio | Mean | StdDev | t-ratio | Mean | StdDev | t-ratio | Mean | StdDev | t-ratio |
| EdgeB | 32.00 | 32.26 | 3.60 | -0.07 | 31.99 | 4.09 | 0.00 | 32.86 | 8.11 | -0.11 | 32.08 | 7.99 | -0.01 |
| Star2B | 114.00 | 115.54 | 22.84 | -0.07 | 114.52 | 26.15 | -0.02 | 118.52 | 50.50 | -0.09 | 115.10 | 49.05 | -0.02 |
| Star3B | 130.00 | 130.77 | 38.78 | -0.02 | 132.35 | 44.41 | -0.05 | 139.14 | 84.05 | -0.11 | 134.97 | 80.45 | -0.06 |
| Star4B | 99.00 | 96.62 | 41.22 | 0.06 | 101.37 | 47.39 | -0.05 | 115.00 | 94.25 | -0.17 | 111.18 | 88.01 | -0.14 |
| Star5B | 50.00 | 46.44 | 28.41 | 0.13 | 51.17 | 33.03 | -0.04 | 69.37 | 78.20 | -0.25 | 66.10 | 69.90 | -0.23 |
| TriangleB | 23.00 | 23.99 | 6.28 | -0.16 | 23.73 | 7.38 | -0.10 | 22.31 | 11.13 | 0.06 | 21.87 | 10.80 | 0.11 |
| Cycle4B | 46.00 | 54.17 | 21.61 | -0.38 | 55.38 | 26.54 | -0.35 | 47.26 | 36.05 | -0.04 | 46.05 | 34.32 | 0.00 |
| ASB | 68.38 | 69.30 | 11.16 | -0.08 | 68.23 | 12.81 | 0.01 | 70.68 | 24.98 | -0.09 | 68.68 | 24.47 | -0.01 |
| ASB2 | 68.38 | 69.30 | 11.16 | -0.08 | 68.23 | 12.81 | 0.01 | 70.68 | 24.98 | -0.09 | 68.68 | 24.47 | -0.01 |
| ATB | 42.50 | 43.08 | 8.03 | -0.07 | 42.43 | 9.18 | 0.01 | 43.39 | 16.92 | -0.05 | 42.48 | 16.57 | 0.00 |
| A2PB | 77.38 | 74.71 | 11.75 | 0.23 | 73.21 | 12.79 | 0.33 | 81.30 | 26.96 | -0.15 | 78.83 | 26.42 | -0.06 |
| AETB | 111.19 | 121.15 | 36.43 | -0.27 | 120.09 | 43.14 | -0.21 | 109.65 | 63.38 | 0.02 | 107.15 | 61.38 | 0.07 |
| Field_MatchB | 15.00 | 14.98 | 2.48 | 0.01 | 15.05 | 2.73 | -0.02 | 15.38 | 4.49 | -0.09 | 16.81 | 4.83 | -0.38 |
| Star2BX | 6149.00 | 6206.22 | 638.51 | -0.09 | 6140.80 | 749.29 | 0.01 | n/a | n/a | n/a | n/a | n/a | n/a |
| StarAB1X | 6904.56 | 6994.06 | 1061.95 | -0.08 | 6890.43 | 1242.04 | 0.01 | n/a | n/a | n/a | n/a | n/a | n/a |
| StarAX1B | 12042.00 | 12154.33 | 1249.03 | -0.09 | 12025.68 | 1466.64 | 0.01 | n/a | n/a | n/a | n/a | n/a | n/a |
| StarAXAB | 3358.00 | 3359.05 | 14.40 | -0.07 | 3357.96 | 16.35 | 0.00 | n/a | n/a | n/a | n/a | n/a | n/a |
| TriangleXBX | 602.00 | 605.83 | 53.17 | -0.07 | 557.67 | 66.65 | 0.67 | n/a | n/a | n/a | n/a | n/a | n/a |
| L3XBX | 296112.00 | 297264.20 | 29363.10 | -0.04 | 293114.23 | 35292.32 | 0.09 | n/a | n/a | n/a | n/a | n/a | n/a |
| ATXBX | 63.70 | 64.21 | 7.14 | -0.07 | 63.67 | 8.12 | 0.00 | n/a | n/a | n/a | n/a | n/a | n/a |
| EXTB | 7155.00 | 7341.72 | 1887.46 | -0.10 | 7245.72 | 2236.19 | -0.04 | n/a | n/a | n/a | n/a | n/a | n/a |
| L3AXB | 27287.00 | 27457.65 | 2678.70 | -0.06 | 27265.83 | 3142.48 | 0.01 | n/a | n/a | n/a | n/a | n/a | n/a |
| C4AXB | 1256.00 | 1268.62 | 122.73 | -0.10 | 1290.55 | 145.39 | -0.24 | n/a | n/a | n/a | n/a | n/a | n/a |
| ASAXASB | 15323.54 | 15413.04 | 1061.95 | -0.08 | 15309.41 | 1242.04 | 0.01 | n/a | n/a | n/a | n/a | n/a | n/a |
| AC4AXB | 10261.00 | 10298.71 | 454.56 | -0.08 | 10253.23 | 531.33 | 0.02 | n/a | n/a | n/a | n/a | n/a | n/a |
| Field_TXBXMatch | 276.00 | 276.83 | 38.65 | -0.02 | 258.40 | 47.43 | 0.37 | n/a | n/a | n/a | n/a | n/a | n/a |
| Field_TXBXMismatch | 326.00 | 328.99 | 40.58 | -0.07 | 299.27 | 49.73 | 0.54 | n/a | n/a | n/a | n/a | n/a | n/a |
| Field_C4AXBMatchB | 540.00 | 544.77 | 88.70 | -0.05 | 611.70 | 105.99 | -0.68 | n/a | n/a | n/a | n/a | n/a | n/a |
| Field_C4AXBMismatch | 716.00 | 723.85 | 93.10 | -0.08 | 678.85 | 110.48 | 0.34 | n/a | n/a | n/a | n/a | n/a | n/a |

**Collaboration models**

| | | *Full socio-cultural* | | | *Socio-cultural* | | | *Full social* | | | *Basic social* | | |
|---|---|---|---|---|---|---|---|---|---|---|---|---|---|
| Statistics | Observed | Mean | StdDev | t-ratio | Mean | StdDev | t-ratio | Mean | StdDev | t-ratio | Mean | StdDev | t-ratio |
| EdgeB | 44.00 | 43.97 | 5.13 | 0.01 | 44.02 | 5.87 | 0.00 | 44.04 | 5.60 | -0.01 | 44.50 | 5.96 | -0.08 |
| Star2B | 192.00 | 187.39 | 39.31 | 0.12 | 189.05 | 45.93 | 0.06 | 189.27 | 43.36 | 0.06 | 192.32 | 46.47 | -0.01 |
| Star3B | 263.00 | 244.88 | 76.11 | 0.24 | 252.61 | 89.62 | 0.12 | 254.99 | 85.02 | 0.09 | 258.07 | 89.80 | 0.06 |
| Star4B | 247.00 | 217.88 | 94.84 | 0.31 | 232.30 | 112.39 | 0.13 | 238.41 | 109.83 | 0.08 | 238.60 | 112.05 | 0.08 |
| Star5B | 171.00 | 138.08 | 84.68 | 0.39 | 153.13 | 100.65 | 0.18 | 161.19 | 103.30 | 0.10 | 158.91 | 101.64 | 0.12 |
| TriangleB | 46.00 | 40.60 | 9.05 | 0.60 | 40.87 | 10.96 | 0.47 | 40.64 | 10.31 | 0.52 | 41.28 | 11.10 | 0.43 |
| Cycle4B | 113.00 | 98.20 | 35.97 | 0.41 | 99.90 | 43.08 | 0.30 | 99.82 | 41.13 | 0.32 | 100.40 | 43.36 | 0.29 |
| ASB | 105.57 | 105.53 | 17.30 | 0.00 | 105.55 | 20.08 | 0.00 | 105.41 | 19.10 | 0.01 | 107.15 | 20.44 | -0.08 |
| ASB2 | 105.57 | 105.53 | 17.30 | 0.00 | 105.55 | 20.08 | 0.00 | 105.41 | 19.10 | 0.01 | 107.15 | 20.44 | -0.08 |
| ATB | 70.25 | 70.24 | 11.09 | 0.00 | 70.15 | 13.24 | 0.01 | 69.93 | 12.50 | 0.03 | 71.13 | 13.44 | -0.07 |
| A2PB | 114.06 | 113.79 | 18.35 | 0.02 | 114.54 | 20.12 | -0.02 | 114.88 | 19.55 | -0.04 | 117.02 | 20.21 | -0.15 |
| AETB | 243.86 | 210.60 | 54.97 | 0.61 | 212.16 | 66.12 | 0.48 | 210.83 | 62.22 | 0.53 | 214.54 | 67.17 | 0.44 |
| Field_MatchB | 30.00 | 30.00 | 4.17 | 0.00 | 29.98 | 4.00 | 0.01 | 29.90 | 4.06 | 0.03 | 22.85 | 3.80 | 1.88 |
| Star2BX | 7245.00 | 7230.36 | 838.76 | 0.02 | 7243.01 | 1020.96 | 0.00 | n/a | n/a | n/a | n/a | n/a | n/a |
| StarAB1X | 8708.82 | 8585.14 | 1424.83 | 0.09 | 8651.08 | 1744.27 | 0.03 | n/a | n/a | n/a | n/a | n/a | n/a |
| StarAX1B | 14138.00 | 14108.94 | 1638.56 | 0.02 | 14133.87 | 1996.71 | 0.00 | n/a | n/a | n/a | n/a | n/a | n/a |
| StarAXAB | 3406.00 | 3405.89 | 20.52 | 0.01 | 3406.07 | 23.49 | 0.00 | n/a | n/a | n/a | n/a | n/a | n/a |
| TriangleXBX | 583.00 | 579.60 | 71.34 | 0.05 | 546.07 | 88.83 | 0.42 | n/a | n/a | n/a | n/a | n/a | n/a |
| L3XBX | 308807.00 | 304876.06 | 37620.77 | 0.10 | 301085.54 | 46946.19 | 0.16 | n/a | n/a | n/a | n/a | n/a | n/a |
| ATXBX | 83.54 | 83.47 | 9.66 | 0.01 | 83.52 | 11.35 | 0.00 | n/a | n/a | n/a | n/a | n/a | n/a |
| EXTB | 10370.00 | 9670.71 | 2124.61 | 0.33 | 9962.25 | 2765.87 | 0.15 | n/a | n/a | n/a | n/a | n/a | n/a |
| L3AXB | 29155.00 | 29256.54 | 3488.46 | -0.03 | 29450.82 | 4367.14 | -0.07 | n/a | n/a | n/a | n/a | n/a | n/a |
| C4AXB | 1167.00 | 1155.85 | 161.11 | 0.07 | 1161.84 | 207.61 | 0.03 | n/a | n/a | n/a | n/a | n/a | n/a |
| ASAXASB | 17127.80 | 17004.12 | 1424.83 | 0.09 | 17070.06 | 1744.27 | 0.03 | n/a | n/a | n/a | n/a | n/a | n/a |
| AC4AXB | 11342.31 | 11340.44 | 591.31 | 0.00 | 11343.37 | 700.36 | 0.00 | n/a | n/a | n/a | n/a | n/a | n/a |
| Field_TXBXMatch | 286.00 | 286.17 | 44.14 | 0.00 | 249.02 | 54.14 | 0.68 | n/a | n/a | n/a | n/a | n/a | n/a |
| Field_TXBXMismatch | 297.00 | 293.43 | 54.19 | 0.07 | 297.06 | 58.52 | 0.00 | n/a | n/a | n/a | n/a | n/a | n/a |
| Field_C4AXBMatchB | 497.00 | 493.16 | 98.69 | 0.04 | 519.94 | 128.84 | -0.18 | n/a | n/a | n/a | n/a | n/a | n/a |
| Field_C4AXBMismatch | 670.00 | 662.69 | 122.83 | 0.06 | 641.90 | 133.04 | 0.21 | n/a | n/a | n/a | n/a | n/a | n/a |